\documentclass[journal]{IEEEtran}
\usepackage{amsmath,amssymb}
  \usepackage[pdftex]{graphicx} 
\begin{document}

\title{Producing Acoustic `Frozen Waves':\\ Simulated experiments}
\author{Jos\'{e}~L.~Prego, Michel~Zamboni-Rached, Erasmo~Recami, and~Hugo~E.~Hern\'{a}ndez-Figueroa

\thanks{
The authors are with the Faculty of Electrical Engineering, State University of Campinas, Campinas, SP,
Brazil (e-mail: mzamboni@dmo.fee.unicamp.br), and E.Recami is also with INFN-Sezione di Milano, Milan, Italy, and
with the Faculty of Engineering of the Bergamo state University, Bergamo, Italy.

This work was mainly supported by FAPESP, CNPq (Brazil), and by INFN (Italy).}
}

\markboth{ }{Prego \MakeLowercase{{\rm et al.}}: Acoustic `Frozen Waves'}

\maketitle

{
\begin{abstract}
{In this paper we show how appropriate superpositions of Bessel beams can be successfully used to obtain arbitrary
longitudinal intensity patterns of nondiffracting ultrasonic wavefields with very high transverse localization.\\
More precisely, the method here described allows generating longitudinal acoustic pressure fields, whose longitudinal
intensity patterns can assume, in principle, any desired shape within a freely chosen interval $0 \leq z \leq L$ of
the propagation axis, and that can be endowed in particular with a {\em static} envelope (within which only the carrier
wave propagates).\\
Indeed, it is here demonstrated by computer evaluations that these very special beams of non-attenuated ultrasonic
field can be generated in water-like media by means of annular transducers. Such fields ``at rest" have been called by us
{\em Acoustic Frozen Waves} (FW).\\
The paper presents various cases of FWs in water, and investigates their aperture characteristics, such as minimum
required size and ring dimensioning, as well as the influence they have on the proper generation of the desired
FW patterns.\\
The FWs are particular {\em localized} solutions to the wave equation that can be used in many applications, like new kinds
of devices, such as, e.g., acoustic tweezers or scalpels, and especially various ultrasound medical apparatus.}\\

\end{abstract}
}


\section{A Preliminary Introduction}
\IEEEPARstart{T}{he} theory of Localized Waves (LW), also called Non-Diffracting Waves, has been developed, experimentally verified
and generalized over the years. The application sectors may include diverse areas such as Optics, Acoustics and Geophysics. One
of their striking properties is that their peak-velocity can assume any value between zero and infinity\cite{hugo2008}.
The most important characteristic of the LWs, however, is that they are solutions to the wave equations capable of resisting the
effects of diffraction, at least up to a certain, long field-depth $L$. They exist both as localized beams and as localized
pulses. \ The LWs that have been more intensely investigated are the so-called ``superluminal" ones\cite{recami2009}, they being the easiest
to be mathematically constructed in closed form, and therefore experimentally generated. Let us recall that the first LWs
of this kind have been mathematically and experimentally created by J.-y. Lu et al. in Acoustics\cite{lu1992,lu1992b}
(when they are actually supersonic, and not superluminal), and used by J.-y. Lu et al. for the construction of a high-definition
ultrasound scanner, directly yielding a 3D image\cite{lu1994}.

  More recently, also for the {\em subluminal} LWs various analytic exact solutions started to be found\cite{zamboni2008}.
Among the subluminal localized waves, the most interesting appear to be those corresponding to zero peak-velocity, that is,
those with a {\em static} envelope (within which only the carrier wave propagates): Such LWs ``at rest" have been called
{\em Frozen Waves} (FW) by us\cite{zamboni2004,zamboni2005,zamboni2006}. The FWs have been actually produced for the first
time, quite recently, in the sector of Optics\cite{tarcio2011}.

  The above information confirms that LW pulses and beams can be used also in ultrasound imaging; but this is not as evident
for the FWs because of the  {\em static} nature of their envelopes. What FWs make possible, by contrast, is the localization
of arbitrary spots of energy within a selected space interval $0 \leq z \leq L$.

  Let us recall here some previous work done in connection with ultrasonic non-diffracting fields. This list,
obviously, is not at all exhaustive, but is merely illustrative of previous literature related with
pulses\cite{lu1992,lu1992b,lu1990,castellanos2010}, with single Bessel beams\cite{hsu1989,holm1998,nowack2012}, with
characteristics of the field emitted by flat annular arrays\cite{fox2002a,fox2002b}, and with scattering produced by
spherical objects\cite{mitri2009,mitri2010,mitri2011}. Many references therein could also be considered. \ Afterwards,
a lot of work has been produced
by investigating all the possible superpositions of Bessel beams obtained via integrations over their axicon angles
(that is, their speed) and/or frequencies and/or wavenumbers and/or phases, etc.: See, e.g., \cite{hugo2008,recami2009}
and the references quoted therein.

  However, only few publications\cite{lu1997,zamboni2011} have addressed the superposition of Bessel
beams with the same frequency but with different longitudinal wavenumbers: In particular for obtaining Frozen
Waves\cite{zamboni2004,zamboni2005,zamboni2006}.

  Purpose of this paper is to contribute to the last topic, by the application of a general procedure previously developed
by us for the generation of FWs (mainly in Optics), and then simulate the production of ultrasonic FW fields in water.
Namely, we shall use the methodology in Refs.\cite{zamboni2004,zamboni2005,zamboni2006},
that allows controlling the longitudinal intensity shape of the resulting fields, confined, as we were saying, inside a
pre-selected space interval
$0 \leq z \leq L$; where $z$ is the propagation axis, while $L$ can be much larger than the wavelength $\lambda$ of the
adopted ultrasonic monochromatic excitation. In practice, we shall perform appropriate superpositions of zero-order $(m=0)$
Bessel beams. The generated fields, apart from possessing a high transverse
localization, are endowed with the important characteristic that inside the chosen interval they can assume any desired shape:
For instance, one or more high-intensity peaks, with distances between them much larger than $\lambda$.\\

  Before going on, let us add a couple of preliminary {\em comments} about the FWs. The first is that it would be of course possible
to use higher-order $(m\geqslant1)$ Bessel beams. In this case, however, it would be practically necessary a subdivision
of the radiator ring elements into small arc segments (i.e., a {\em segmented array\/}\cite{akhnak2002}), because of the
azimuthal phase dependence $e^{im\phi}$ of
those fields on the beam axis $z$. It appears to be more convenient tackling with such complexities in the aperture design only
after that an investigation of the cylindrically symmetric zero-order Bessel beam superpositions has been exploited. \
Anyway, we shall briefly come back, in Sec.2, to the higher-order Bessed beams question. \
The second comment is related to the flux of energy within a FW, especially in the realistic case of a finite aperture.
Let us go back for a moment to a (truncated) Bessel beam, by recalling first of all that its good properties are due
to the fact that, even in presence of diffraction, the ``intensity rings" that are known to constitute its transverse
structure (and whose values decrease when the spatial transverse coordinate $\rho$ increases) go on reconstructing the beam
itself all along a (large) depth of field. Namely, given a Bessel beam and a gaussian beam ---both with the same energy $E$,
the same spot $\Delta\rho_0$ and produced by apertures with the same redius $R$ in the plane $z=0$---, the {\em percentage}
of energy $E$ contained in the central peak region $0 \leq \rho \leq \Delta\rho_0$ is {\em smaller} for a Bessel rather than
for a gaussian beam: It is just such a different distribution of energy on the trasnsverse plane that causes the
{\em reconstruction} of the Bessel beam central peak even at large distances from the source, and even after an obstacle
with sizes smaller than the {\em aperture\/}'s. Such a property is possessed also by the other localized waves. \ In
other words, a certain energy {\em must} be contained in, and carried by, the side-lobes!  Of course, such energy can be
reduced towards its minimum necessary value by suitable techniques: See, e.g., Refs.\cite{lu11,lu12}.

  Also the FWs we are dealing with in this paper are non-diffracting beams: and we investigate their generation by a limited
aperture, so that their energy flux does remain finite. But they are just the energy contributions coming from the lateral
rings that strengthen the FW field pattern, in the $z$ direction, all along the FW field depth.

%

  The paper is organized as follows: After this Introduction, Section 2 briefly describes the
methodology used for the generation of FWs. \ Then, in Section 3 the method for the computation of the ultrasonic fields
is presented; followed in Section 4 by a discussion about the annular radiator requirements. \
With respect to the last point, let us observe that, even if the precision requirements that one meets for the annular
transducer sizes are tight, and appear as a real challenge to be overcome, nevertheless it is encouraging to know that
various successes have been already obtained in similar situations, for example for monolithic\cite{lu1990},
piezocomposite\cite{akhnak2002}, and PVDF transducers\cite{Piezoflex}. More explicitly, let us mention that
ultrasonic beams have been technologically synthesized, by superposing Bessel radiations, in interesting works
like \cite{domell1982,foster1989,hsu1989,lu1994b,eiras2003,aulet2006,moreno2010,calas2010,castellanos2011}.

\ Finally, four simulated examples of ultrasonic FW fields in an ideal water-like
medium assuming no attenuation are presented in Section 5; while some conclusions appear at the end of the paper.

\section{Introduction to the Frozen Waves (FW)}

\noindent In this Section, a method for the generation of FWs is presented. For more details the reader is referred to
reference\cite{zamboni2005}. The main aim here is constructing, inside the finite interval $0\leq z \leq L$
of the propagation axis ($\rho=0$), a {\em stationary} intensity envelope with a desired shape, that we call $|F(z)|^2$.
To such a purpose, we shall take advantage of the field localization features of the axis-symmetrical zero-order
Bessel beams. \ Of course, we could start from higher order Bessel beams, and their relevant formalism.  But, for simplicity,
as already said, we prefere to start from the FWs given by the following finite superposition of zero-order Bessel function
of the $1^{st}$ kind, all with the same frequency but with different (and still unknown) longitudinal wavenumbers $\beta_n \,$:
\begin{equation}
\label{eqn_1}
  \Psi(\rho,z,t) = e^{-i\omega_0t} \sum\limits_{n=-N}^{N} A_n
  J_0(k_{\rho n}\rho) e^{i\beta_n z} \ .
\end{equation}

\noindent In Eq.(\ref{eqn_1}), quantities $k_{\rho n}$ are the transverse wave numbers of the Bessel beams, linked to the
values of $\beta_n$ by the following relationship:
\begin{equation}
\label{eqn_2}
  k_{\rho n}^2 + \beta_{n}^2 = \frac{\omega_0^2}{c^2} \ ,
\end{equation}

\noindent where $\omega_0=2\pi f_0$, is the angular frequency, and $c$ the plane wave phase velocity in the selected medium.
By the way, let us recall that we leave the frequency fixed, since we are considering beams (and not pulses): We vary,
afterward, the longitudinal wavenumber (and amplitude and phase) of each Bessel beam, so as to obtain the desired longitudinal
shape of the resulting beam.

  It is important to notice that we restrict the values of the longitudinal wave numbers to the interval
\begin{equation}
\label{eqn_3}
0 \leq \beta_n \leq \frac{\omega_0}{c} \; ,
\end{equation}

\noindent in order to avoid evanescent waves and to ensure forward propagation only.

  Our goal is now finding out the values of the $\beta_n$, and of the constant coefficients $A_n$ in Eq.\ref{eqn_1},
in order to reproduce approximately, inside the said interval $0\leq z \leq L$, the desired longitudinal intensity
pattern $|F(z)|^2$. \ In other words, for $\rho=0$ we need to have:
\begin{equation}
\label{eqn_4}
  \sum\limits_{n=-N}^{N} A_n e^{i\beta_n z} \approx F(z) \ \ \ \ \text{with} \ 0\leq z \leq L \; .
\end{equation}

\noindent To obtain this, one possibility is to take $\beta_n=\frac{2\pi n}{L}$, thus obtaining a truncated Fourier series
that represents the desired pattern $F(z)$. This choice, however, is not very appropriate because of two
reasons: \ (i) it yields negative values for $\beta_n$ when $n<0$; which would imply backward propagations; \ (ii)
we wish to have $L\gg \lambda$, and in this case the main terms in the series could lead to very small values of the $\beta_n$,
resulting in a very short depth of field, and strongly affecting, therefore, the generation of the desired envelopes
far form the source. \ A possible way out is to take
\begin{equation}
\label{eqn_5}
  \beta_{n} = Q + \frac{2\pi}{L}n \ ,
\end{equation}

\noindent where the value of $Q>0$ can be freely selected. Then, on introducing Eq.(\ref{eqn_5}) into Eq.(\ref{eqn_2}),
the transverse wavenumbers can be expressed as
\begin{equation}
\label{eqn_6}
  k_{\rho n}^2 = \frac{\omega_0^2}{c^2} - \left(Q + \frac{2\pi}{L}n\right)^2 \; .
\end{equation}

\noindent Now, inserting again Eq.(\ref{eqn_5}) for $\beta_n$ into (\ref{eqn_1}) and putting $\rho=0$, one gets
\begin{equation}
\label{eqn_7}
  \Psi(\rho=0,z,t) = e^{-i\omega_0t} e^{iQz}\sum\limits_{n=-N}^{N} A_n\;e^{i\frac{2\pi n}{L}z} \ ,
\end{equation}

\noindent with
\begin{equation}
\label{eqn_8}
  A_n = \frac{1}{L} \int_{0}^{L} F(z)\;e^{-i\frac{2\pi n}{L}z} dz \ .
\end{equation}

\noindent Expressions (\ref{eqn_7}) and (\ref{eqn_8}) represent an approximation of the desired longitudinal pattern, since the
superposition in (\ref{eqn_1}) is necessarily truncated. However, we can control, and improve, the fidelity of the reconstruction
by varying the total number of terms $2N+1$ in the finite superposition by a suitable choice of the parameters $Q$ and/or
$L$. \ The complex coefficients $A_n$ in (\ref{eqn_8}) forward the final amplitudes and phases for each one of the Bessel
beams in Eq.(\ref{eqn_1}); and, because we are adding together zero-order Bessel functions, we can expect a high
degree of field concentration around $\rho=0$.

  The methodology introduced here deals with control over the longitudinal intensity pattern. Obviously, we cannot get a total
3D control, owing to the fact that the field must obey the wave equation. However, we can get some control over the
transverse spot size through the parameter $Q$. \ Actually, Eq.(\ref{eqn_1}), which defines our FW, is a superposition of
zero-order Bessel beams, and therefore we expect that the resulting field possesses an important transverse localization
around $\rho=0$. Each Bessel beam in superposition (\ref{eqn_1}) is associated with a central spot with transverse width
$\Delta\rho_n \approx 2.4/k_{\rho n}$; so that, on the basis of the expected convergence of series (\ref{eqn_1}), we can
estimate the radius of the transverse spot of the resulting beam as being:
\begin{equation}
\label{eqn_9}
\Delta \rho \approx \frac{2.4}{k_{\rho,n=0}} = \frac{2.4}{\sqrt{\omega_0^2/c^2 - Q^2}} \ .
\end{equation}

  Let us explicitly notice, at this point, that one can increase the control over the transverse intensity pattern
of the resulting beam by having recourse to the already mentioned higher order Bessel beams, \ $A_n \exp(-i\omega t)
J_{\nu}(k_{\rho\,n}\rho)\exp(i \beta_n z)$, \ with $\nu \geq 1$, \ in the superposition (\ref{eqn_1}),
keeping the same values of the $A_n$ given by Eq.(\ref{eqn_8}), and of the $\beta_n$ given by Eq.(\ref{eqn_5}). \
On doing this, the longitudinal intensity pattern can be shifted from $\rho = 0$ to a cylindrical surface
of radius $\rho = \rho'$, which can be approximately evaluated through the relation

$$(\frac{d}{d\rho}J_{\nu}(\rho \sqrt{\omega^2/c^2 - Q^2}))|_{\rho=\rho'} = 0 \; .$$

\noindent This allows one to get interesting nondiffracting fields, modelled over cylindrical surfaces. Some details can
be found in Ref.\cite{zamboni2005}.

   Anyway, relationship (\ref{eqn_9}) itself is rather useful, because, once we have chosen the desired
longitudinal intensity pattern $|F(z)|^2$, we can also choose the size of the transverse spot ($2\Delta\rho$),
and then use Eq.(\ref{eqn_9}) to evaluate the corresponding, needed value of $Q$.

  Such a method for creating FWs depends on the nature of the waves to be used. For instance, for an efficient generation in
Optics, a recourse to ordinary lenses is required even when using the classical method with an array of so-called Durnin et al.'s
circular slits\cite{sheppard1,sheppard2,durnin1987}.
In several other cases, optical axicon lenses have been used\cite{McLeod1,McLeod2}
even if, to say the truth, axicon devices had been already proposed since long time\cite{UltrasAxic}
in Acoustics too.  Of course, whenever possible, one may have recourse also to holographic elements, or suitable mirrors, etc. \
For instance, optical FWs were produced experimentally (for the first time) by using computer generated holograms and a
spatial light modulator\cite{tarcio2011}.

  In the case of ultrasound, we may adopt annular transducers composed of many narrow rings. Each one of them will
have to be exited by a sinusoidal input having a particular amplitude and phase. \ Section 4 will discuss these points
in more detail.

\section{Method for Calculating Ultrasonic Fields}

\noindent This Section summarizes the well known impulse response (IR) method, which is the mathematical
basis used in this work for the computation of the ultrasonic frozen waves. \ Such a spatial impulse response method
has recourse to the linear system theory to separate the temporal from the spatial characteristics of the acoustic field. \
Then, the aperture IR function can be derived from the Rayleigh-Sommerfeld formulation of diffraction,
that is, from the Rayleigh integral\cite{stepanishen1971,harris1981,jensen1992,goodman2005}, via the
following expression:
\begin{equation}
\label{eqn_10}
  h(\mathbf{r_1},t) = \int_{\mathcal{S}}
  \frac{\delta\left(t-\frac{|\mathbf{r_1-r_0}|}{c}\right)}{2\pi |\mathbf{r_1-r_0}|} d\mathcal{S}\ .
\end{equation}

\noindent Here the radiating aperture $S$ is mounted on an infinite rigid baffle; quantity $c$ being now the speed of sound,
and $t$ the time. The spatial position of the field point is designated by $\mathbf{r_1}$, while $\mathbf{r_0}$ denotes
the location of the radiating aperture.

  The integral \ref{eqn_10} is basically the statement of Huyghens principle, and evaluates the acoustic
field, i.e., the field pressure relative to the unperturbed static pressure $P_0$, by adding the spherical wave
contributions from all the small area elements that constitute the aperture.  \ This process can also be reformulated
by using the acoustic reciprocity principle, and then constructing $h$ by finding out the part of the spherical waves
that intersect the aperture.  \ This implies that the IR function $h$ depends both on the form of the radiating element and
on its relative position w.r.t. (with respect to) the point where the acoustic pressure is calculated. \ The IR formulation assumes
a flat or gently curved aperture (that is, the aperture dimensions are large compared to the wavelength), which
radiates into a homogeneous medium with no attenuation,\footnote{It is possible to include the attenuation of the medium by
filtering the Fourier transform of $h$ by a frequency dependent attenuation function\cite{harris1981,jensen1993}.} and operates in
a linear regime.

  On using the spatial impulse response $h(x,y,z,t)$, the final acoustic pressure can be written as\cite{harris1981,jensen1993}:
\begin{equation} \label{eqn_11}
p(x,y,z,t) = \rho_{m} \frac{\partial v(t)}{\partial t} * h(x,y,z,t) \ ,
\end{equation}

\noindent where $\rho_{m}$ is the density of the medium, and the symbol star, $*$, denotes the convolution in time of the
derivative of the velocity surface signal $v$, for a radiating element with aperture's IR function $h$.

  The above procedure was implemented into a matlab program, employing the DREAM toolbox\cite{piwakowsky1989,piwakowsky1999}
for the calculation of the IR of the rings that compose the annular aperture. The steps followed by the program for computing
the acoustic fields can be summarized as follows:

\begin{enumerate}
\item Compute the theoretical FW desired pattern.
\item Determine the transducer requirements.
\item Sample the amplitude \& phase FW profiles at $z=0$.
\item Assign sinusoidal $v_k$ for ring $n_k$ using sampled values.
\item Sweep field points $x_{i,j};y_{i,j};z_{i,j}$ for transducer ring $n_k$.
\item At point $P_{(i,j)}$ calculate $h$ and $p_{(i,j)} = \rho_{m} \dot{v}_k*h$.
\item Accumulate pressure for ring $n_k$ \& go back to step $4$.
\item Store and display results.
\end{enumerate}

\noindent The medium selected for the simulations was an ideal water-like lossless medium,\footnote{The case
including attenuation will be investigated in a forthcoming paper.} having a constant propagation
speed of $c=1540 \;$m/s [which is an average\cite{rossing2007} among  $c_{{\rm fat}}=1450\;$m/s, \ $c_{{\rm blood}}=1575\;$m/s,
and $c_{{\rm muscle}}=1600\;$m/s].

\section{Aperture Dimensioning}

\noindent In this Section we discuss some dimensioning aspects of an annular aperture for creating the FW fields.
They include: \ i) the selection of the transducer size; and \ ii) the number and dimensions of the rings.

\subsection{Transducer size}
\noindent The first element to be considered is the minimum aperture radius, $(R_ {{\rm min}})$, required for the generation of the
desired ultrasonic FW fields. \ The selection of this value, however, is not entirely free, because it depends on the chosen
$N$, that determines the number ($2N+1$) of Bessel beams added in Eq.(\ref{eqn_1}). \ From Eqs.(\ref{eqn_3})
and (\ref{eqn_5}) we can write:
\begin{align}\label{eqn_12}
  \beta_ {{\rm max}} &= Q + \frac{2\pi N}{L} \leq \frac{\omega_0}{c} \\
  \beta_ {{\rm min}} &= Q - \frac{2\pi N}{L} \geq 0 \notag  \; .
\end{align}

\noindent Afterwards, if we put $\beta_ {{\rm max}}=\frac{\omega_0}{c}$ into Eqs.(\ref{eqn_12}), and subtracts both sides of the
equations, we get
\begin{equation}\label{eqn_13}
\beta_ {{\rm min}}=\frac{w_0}{c}-\frac{4\pi N}{L} \ .
\end{equation}

\noindent The equation for the maximum axicon-angle is thus
\begin{equation}\label{eqn_14}
  \tan \theta_ {{\rm max}} = \frac{R_ {{\rm min}}}{L} = \frac{k_{\rho_ {{\rm max}}}}{\beta_ {{\rm min}}}\ .
\end{equation}

  Inserting the expression for $k_{\rho}$ from Eq.(\ref{eqn_2}) into Eq.(\ref{eqn_14})
and solving it for $\beta_ {{\rm min}}$, we get:
\begin{equation}\label{eqn_15}
  \beta_ {{\rm min}} = \frac{\frac{w_0}{c}}{\sqrt{1+\frac{R_ {{\rm min}}^2}{L^2}}}\ .
\end{equation}

  Finally, is we equate expressions (\ref{eqn_13} and \ref{eqn_15}), we obtain for $R_ {{\rm min}}$ the expression
\begin{equation}\label{eqn_16}
  R_ {{\rm min}} = L \sqrt{\frac{\frac{w_0^2}{c^2}}{\left(\frac{w_0}{c}-\frac{4\pi N}{L}\right)^2} -1} \ ,
\end{equation}

\begin{figure}[t] 
\centering
\includegraphics[width=3.5in]{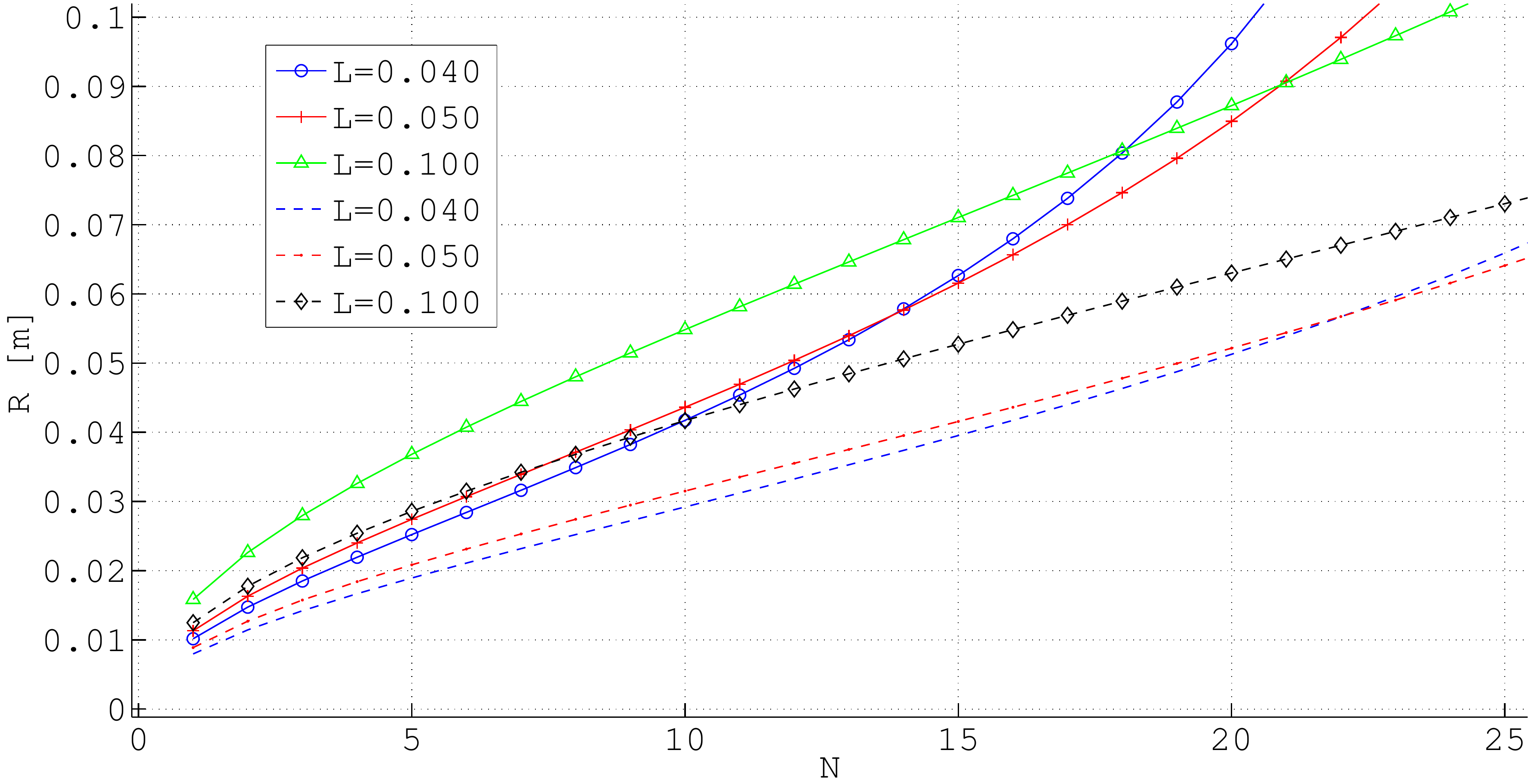}
\caption{Relationship between the aperture radius $R$ and the summation limits $N$ entering Eq.(\ref{eqn_4}), in terms of the
parameters $L$ and $w_0=2\pi f_0$. Continuous lines refer to frequency $f_0=2.5\;$MHz, while dashed lines refer to
$f_0=4\;$MHz.}
\label{fig_1}
\end{figure}

\noindent which does allow the estimation of the minimum aperture-radius for the selected value of $N$, with $\omega_0$ and $L$
as parameters.

  If requisite (\ref{eqn_16}) is not fulfilled, and $R<R_ {{\rm min}}$, the resulting FW pattern may result distorted. This is because
the higher is the precision desired for the FW, or the larger its desired maximum distance $L$, the higher will be the values
needed for $N^\uparrow$ and the aperture radius $R^\uparrow$. This is a logical conclusion, if we think the frozen waves
as nothing but the realization of Huyghens principle for constructive/destructive interference. \ We can also get the above
mentioned dependence by putting $\beta_ {{\rm min}}=0$ in Eq.(\ref{eqn_13}), and then solving it for $N$, so as to obtain
the \emph{maximum} number of allowable terms $N_ {{\rm max}}$ (this implies setting $Q=\frac{w_0}{2c}$) in the Fourier
superposition:

\begin{equation}\label{eqn_17}
  N \leq N_ {{\rm max}} = \frac{L\omega_0}{4\pi c} \ .
\end{equation}

\begin{figure}[t]  
\centering
\includegraphics[width=3.5in]{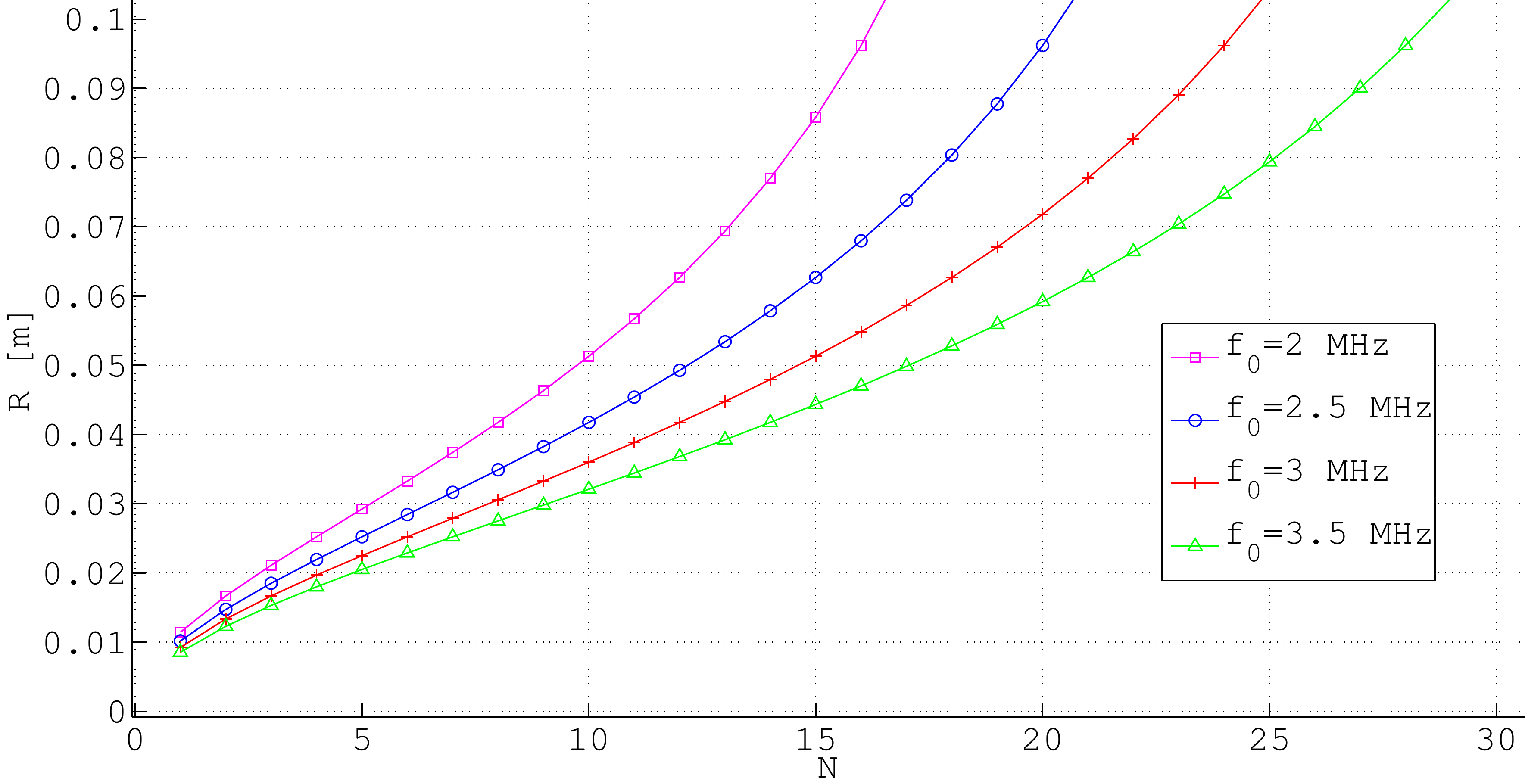}
\caption{Relationship between the aperture radius $R$ and the number $N$ which determines the number of the
Fourier terms entering Eq.(\ref{eqn_4}), for a fixed distance $L=0.040\;$m  but for various values of $f_0$.} \label{fig_2}
\end{figure}

  However, care has to be taken when using using this limiting value, since usually it leads to unpractical transducer
sizes ($R > 50\;$mm), with too many rings ($N_r>100$).

  Relationship (\ref{eqn_16}) is at work in Figures \ref{fig_1} and \ref{fig_2}, where one of the parameters ($L$ or $w_0$)
is alternatively fixed, while the other is left variable.\footnote{We'd like to point out in these Figures that
the estimated size of the spot-radius given by Eq.(\ref{eqn_9}) changes when $N$ is varied. This is because $Q$ disappears
during the derivation of Eq.(\ref{eqn_16}). However, for the current setting, the changes are not significant
and are in the range \ $0.3 \leq \Delta\rho \leq 0.5\;$mm.}  Both Figures clearly show how the value of
$R$ rapidly increases when $N$ is increased. \ By contrast, Figure \ref{fig_1} depicts Eq.(\ref{eqn_16}) for different values
of $L$, using the two frequencies $f_0=2.5\;$MHz, \ and \  $f_0=4\;$MHz; \ and show how in both cases, up to a certain value
of $N$ [e.g. $N\thickapprox 15$ for the first case, continuous lines], the value of $R$ almost does not change when the
distance $L$ is increased.

  The rapid increase of $R$ with $N$ can be partially mitigated by raising the working frequency
$f_0^\uparrow$, as shown in Fig.\ref{fig_2}, were a fixed distance $L=0.040\;$m is assumed. \
This option, however, has the effect of requiring smaller ring-widths ($d^\downarrow$) for a good generation of the
fields.

\begin{figure}[t]  
\centering
\includegraphics[width=3.25in]{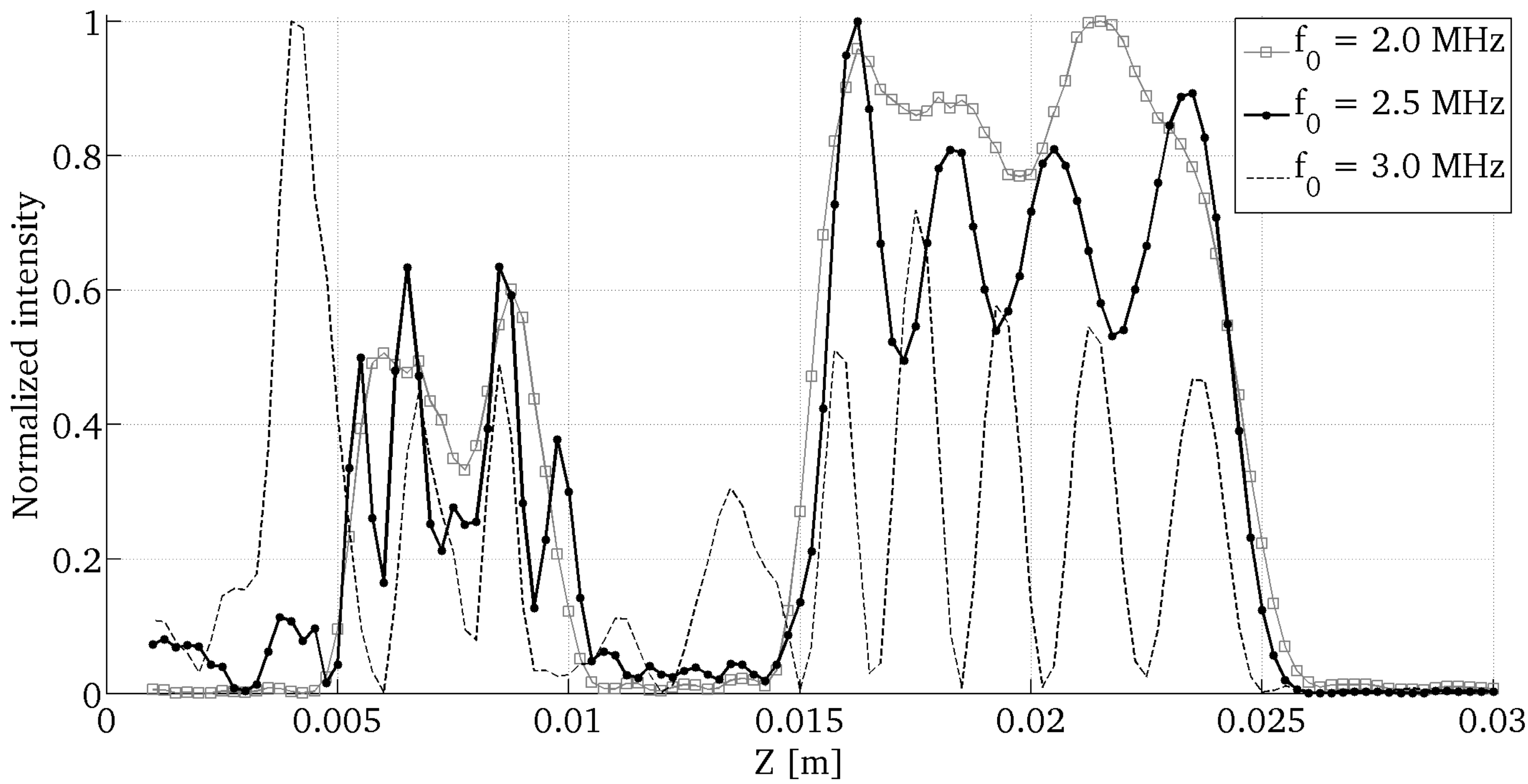}
\caption{Side view of the three {\em simulated} profiles, for a simple (ideal) FW consisting of two step-functions only
[see case Case 1 of Sec.5: \ Cf. Eq.(17) below], obtained by varying the working frequency. The parameter settings
are the following: \ $N=12$; \ $R=35\;$mm; \ $d=0.3\;$mm; \ $\Delta_d=0.05\;$mm; \ $N_r=101$; \ $L=30\;$mm, \ and \ $f_0=2.5\;$MHz}
\label{fig_3}
\end{figure}

  In Fig.\ref{fig_3} we can see how the frequency affects the {\em simulated} profiles, in the simple case of an ideal
FW consisting of two step-functions only [a situation better exploited in Case 1 of Sec.5: Cf. Eq.(17) below]. The
three intensity profiles simulated in Fig.\ref{fig_3} have been obtained by varying the working frequency, always keeping the
same aperture. \ Notice how, for frequencies larger than $f\approx2.5\;$MHz, any increase in the frequency distorts
more and more the envelope of the desired FW. \
This effect is due to two causes: First, the dimensions of the rings are not changed, and , second, the patterns that result
from the beam superpositions in Eq.(\ref{eqn_1}) are influenced by the approximately linear relationship existing between the
ultrasound wavelength and the average distance among the sidelobe peaks of the superposed Bessel functions. \
At the end, the combination of these two causes produce a destructive interference effect on the generated FW.

\begin{figure}[t]  
\vspace{-0.5mm}
\centering
\includegraphics[width=3.25in]{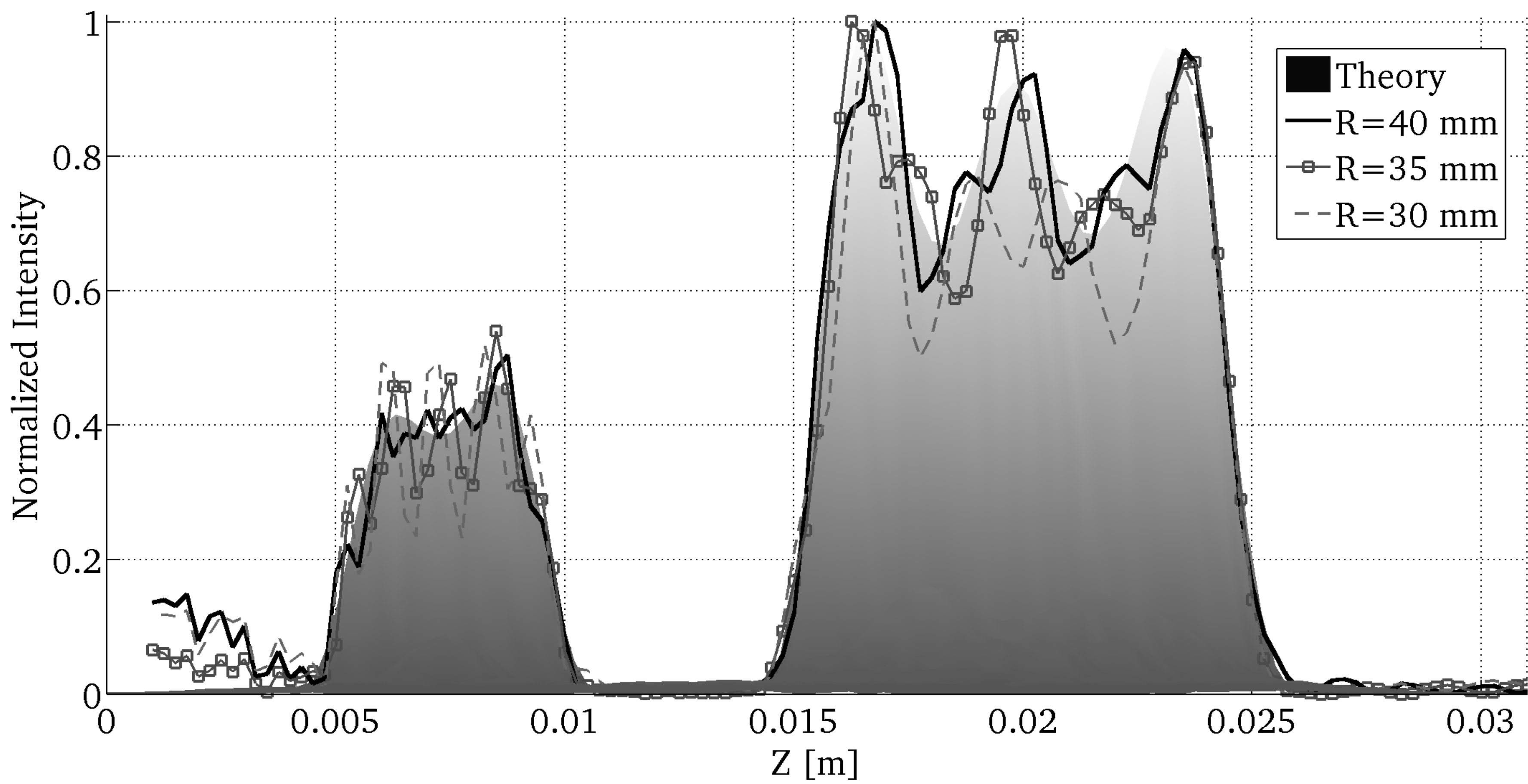}
\caption{Comparison of of the simulated profiles, of the FW considered in Fig.3, obtained now by using three different values
for the transducer radius $R$. \ Settings: $N=9$; \ $d=0.3\;$mm; \ $\Delta_d=0.05\;$mm; \ $L=30\;$mm; \ $f_0=2.5\;$MHz; \
$N_{r1}=115$; \ $N_{r2}=101$, \ and \ $N_{r3}=86$.}
\label{fig_4}
\end{figure}

  The effect of changing the emitter radius $R$ can be observed in Figure \ref{fig_4}, which still refers to the
FW considered in the previous graphic: That is, to an ideal FW consisting of two step-functions
only [a case better exploited, let us repeat, in Figure \ref{fig_8} of Case 1 in Sec.5: Cf. Eq.(17) below]. \
This time, the three simulated intensity profiles have been obtained using the different transducer radii\ $R_1=40\;$mm, \
$R_2=35\;$mm \ and \ $R_3=30\;$mm. \ Other parameters used for the simulation are: \ $L=30\;$mm, \ $d=0.3\;$mm, \
$\Delta_d=0.05\;$mm, \ $N=9$ \ and \ $f_0=2.5\;$MHz. \
Notice how, when the radius is reduced (e.g., $R=30\;$mm), the FW intensity profile gets more distorted with respect
to the ideal envelope.

\begin{figure}[t]  
\vspace{-0.5mm}
\centering
\includegraphics[width=3.25in]{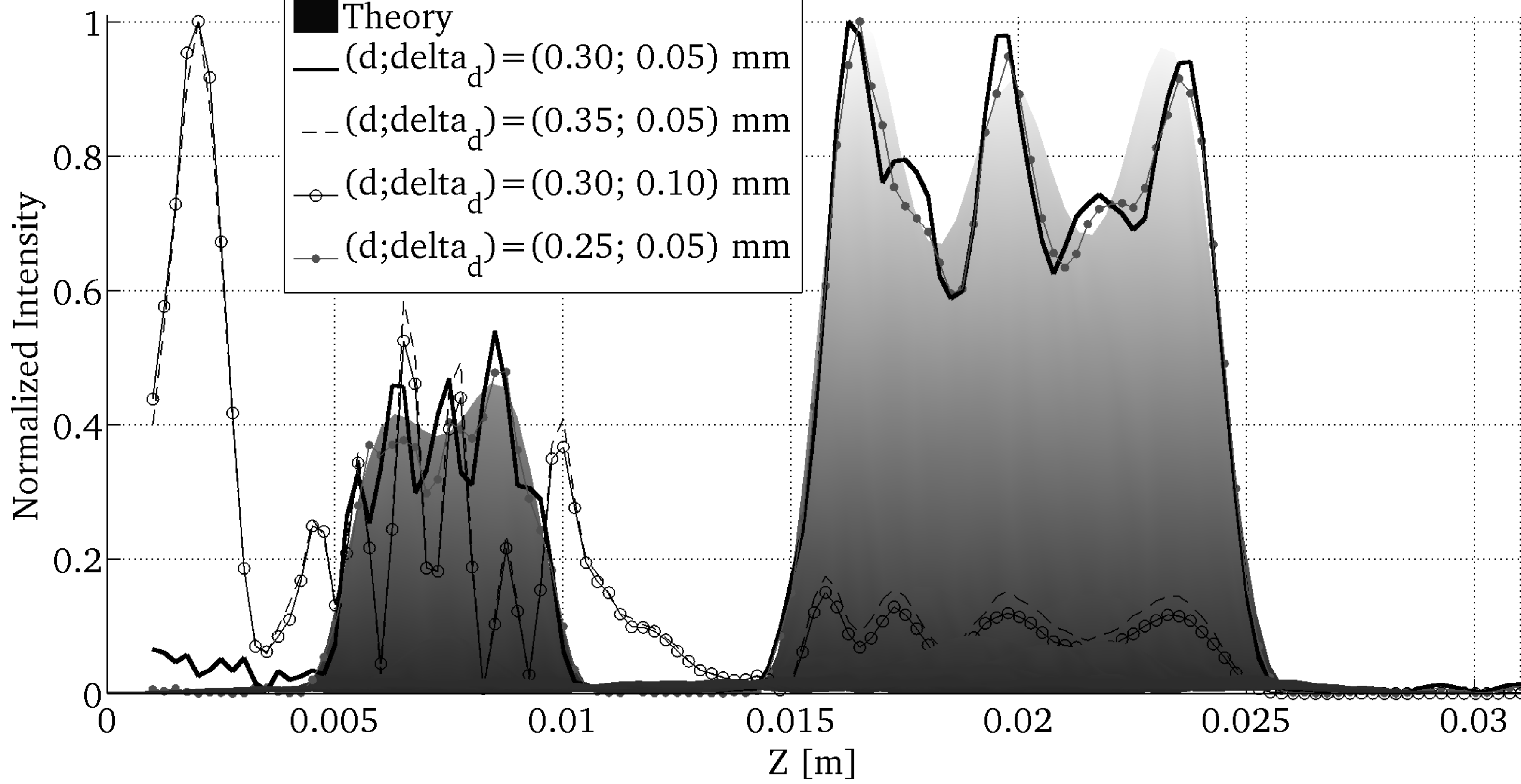}
\caption{Comparison of the simulated FW profiles when using different values for $d; \ \Delta_d$. \ Settings: \
$R=35$\;mm; \ $N=9$; \ $L=30$\;mm; \ $f_0=2.5$\;MHz; \ $N_{r1}=101$; \ $N_{r2}=88$; \ $N_{r3}=88$\ and $N_{r4}=117$.}
\label{fig_5}
\end{figure}

\subsection{Dimensions of the Rings}
\noindent The next step in the definition of the annular radiator is the dimensioning of the transducer rings. This means
determining: \ (i) the width \ $d$, \  and \ (ii) the inter-ring spacing, or kerf, $\Delta_d$.\\
On this respect, one meets a practical problem when the required distances are very small, for instance
sub-millimetric dimensions. In this case, the technology employed in the annular transducer fabrication
may strongly influence its performance. \
To illustrate this effect on the generated ultrasonic fields, in Figure \ref{fig_5} we show how small changes
in $d$ or $\Delta_d$ can lead to severe distortions in the profiles (and the same will be true, e.g., for Fig.8 below).
The corresponding parameters ($d_i; \ \Delta_{d_i}$) chosen for this Figure are: First choice: ($0.30; \ 0.05$) mm; \ Second
choice: ($0.35; \ 0.05$) mm; \ Third choice: ($0.30; \ 0.10$) mm; \ and Fourth choice: ($0.25; \ 0.05$) mm.

\hyphenation{cri-ti-cal}
  One can observe how, for dimensions larger than $d\approx0.30$\;mm and $\Delta_d\approx 0.05$\;mm, the FW envelopes gets
heavily modified. This phenomenon is probably related to the above mentioned causes of destructive interference, as well as
to a near-field effect of the radiator. One can attempt at partially overcome it by increasing the maximum distance
$L^\uparrow$ of the FW fields.

\begin{figure}[t]  
\centering
\includegraphics[width=3.5in]{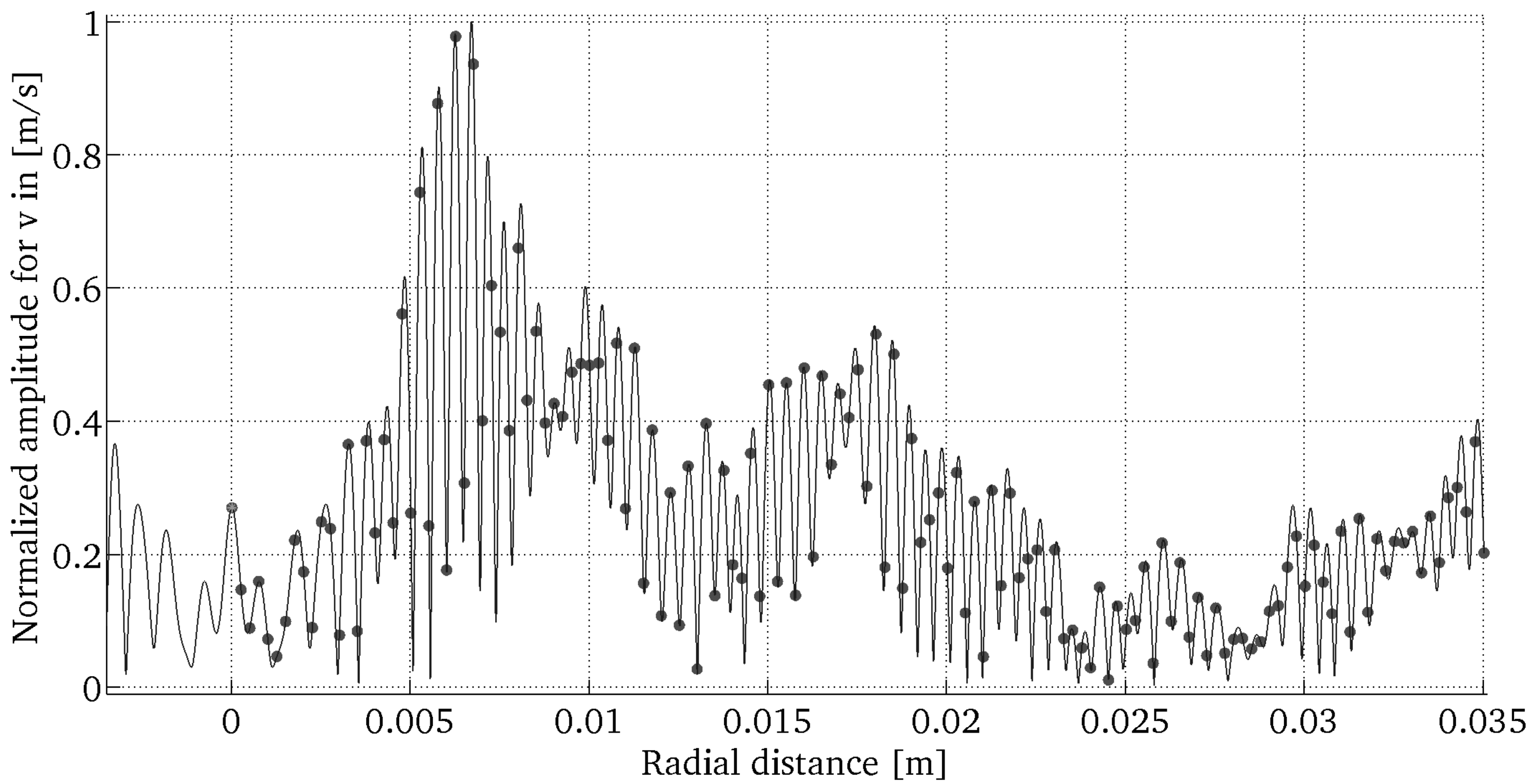}
\caption{Sampling of the ideal symmetric {\em amplitude} pattern, for the generation of the field
in Fig.\ref{fig_8} below. The dots indicate the sampled values used for the excitation of the rings.}
\label{fig_6}
\end{figure}

\begin{figure}[t]  
\centering
\includegraphics[width=3.5in]{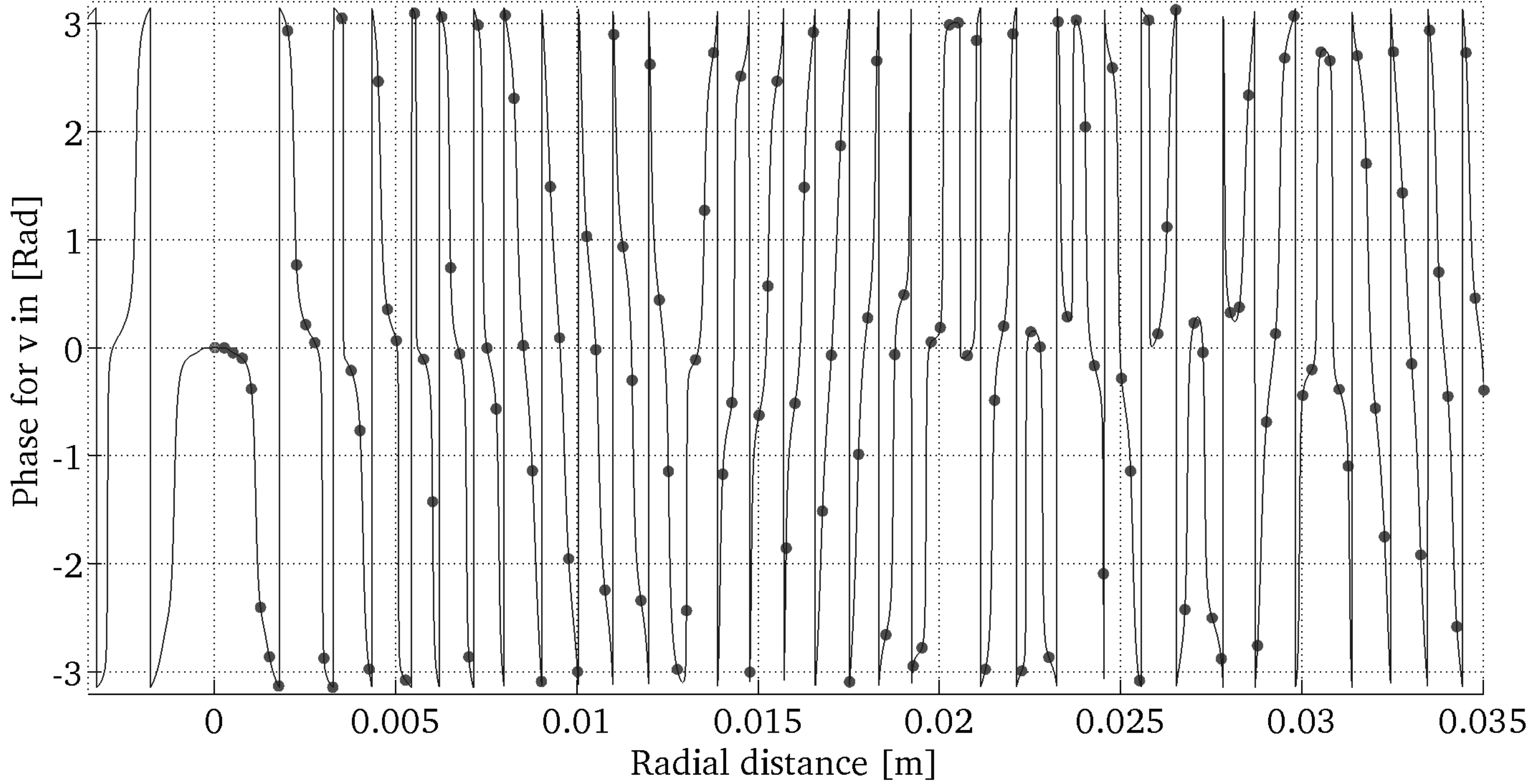}
\caption{Sampling of the ideal symmetric {\em phase} pattern, for the generation of the field in the following Fig.\ref{fig_8}.
The dots indicate the sampled values used for the excitation of the transducer rings.}
\label{fig_7}
\end{figure}

\begin{figure}[t]  
\vspace{-0.5mm}\centering
\includegraphics[width=3.5in]{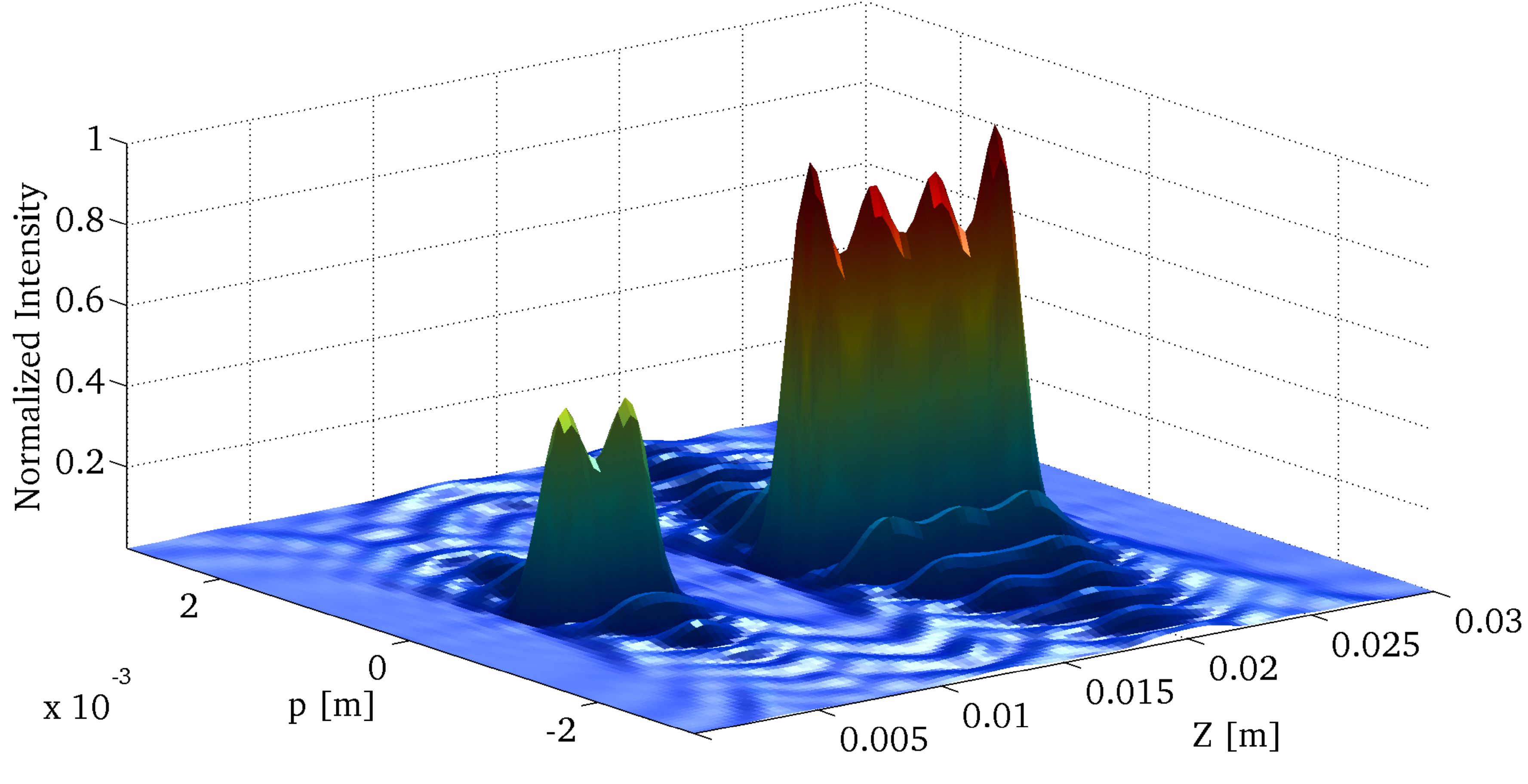}
\caption{The theoretical FW pattern chosen in the Case 1 of Sec.5. \ Settings: $N=12$; \ $R=35\;$mm; \ $d=0.3\;$mm; \
$\Delta_d=0.05\;$mm; \ $N_r=101$;  \ $L=30\;$mm;  \ and \ $f_0=2.5\;$MHz.}
\label{fig_8}
\end{figure}

\begin{figure}[t]  
\vspace{-0.5mm}\centering
\includegraphics[width=3.5in]{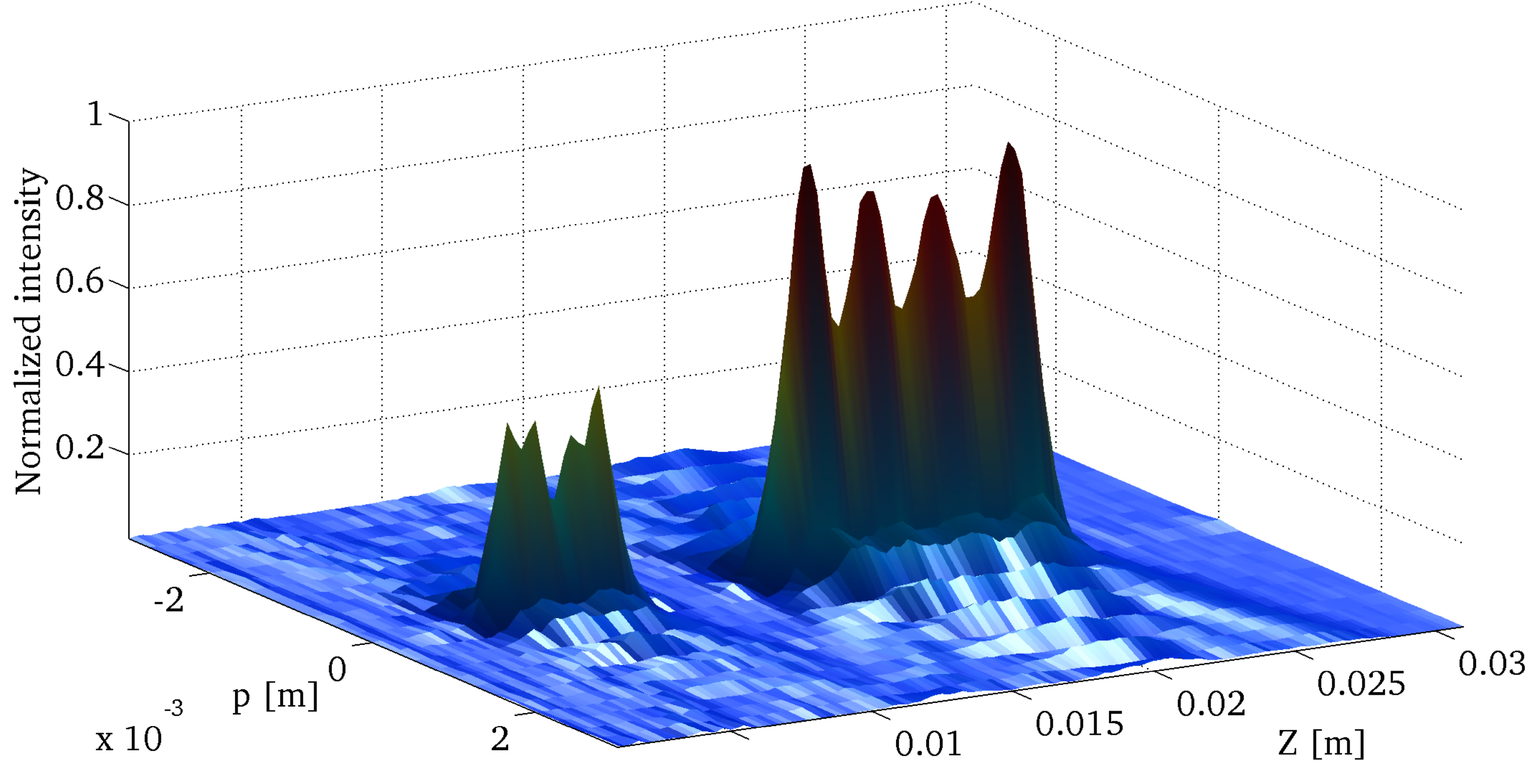}
\caption{The FW pattern, corresponding to the previous Figure, that is, to Case 1 of Sec.5, obtained by our simulated experiment.
The parameters are set as in Fig.3: Namely, \ $N=12$; \ $R=35\;$mm; \ $d=0.3\;$mm; \ $\Delta_d=0.05\;$mm; \ $N_r=101$; \
$L=30\;$mm \ and \ $f_0=2.5\;$MHz.}
\label{fig_9}
\end{figure}

\begin{figure}[t]  
\centering
\vspace{-2.5mm}
\includegraphics[width=3.5in]{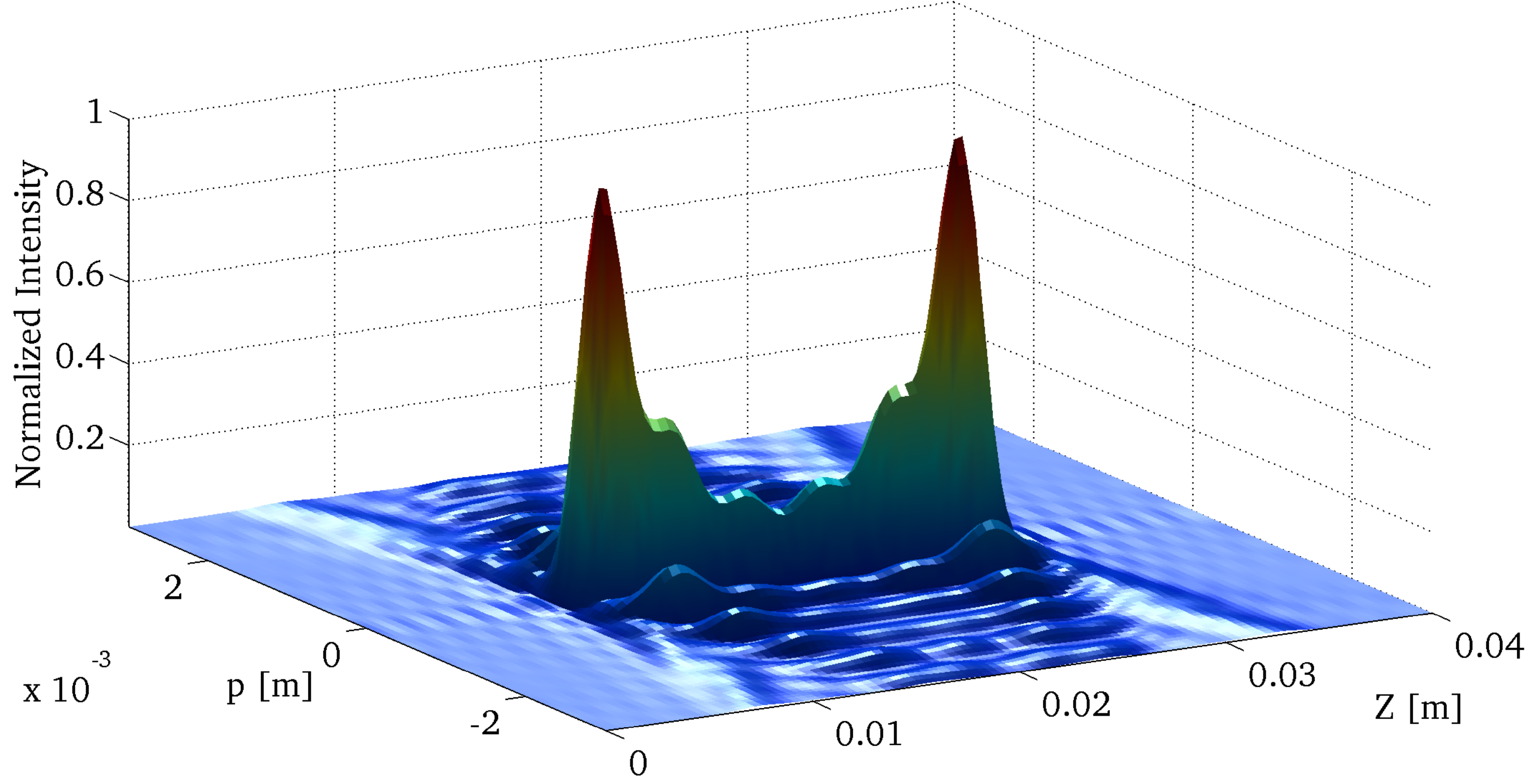}
\caption{The theoretical FW pattern chosen in Case 2. \ Settings: $N=12$; \ $R=35\;$mm; \ $d=0.3\;$mm; \ $\Delta_d=0.05\;$mm; \
$N_r=101$; \ $L=40\;$mm; \ and \ $f_0=2.5\;$MHz.}
\label{fig_10}
\end{figure}

\begin{figure}[t]  
\centering
\vspace{-2.5mm}
\includegraphics[width=3.5in]{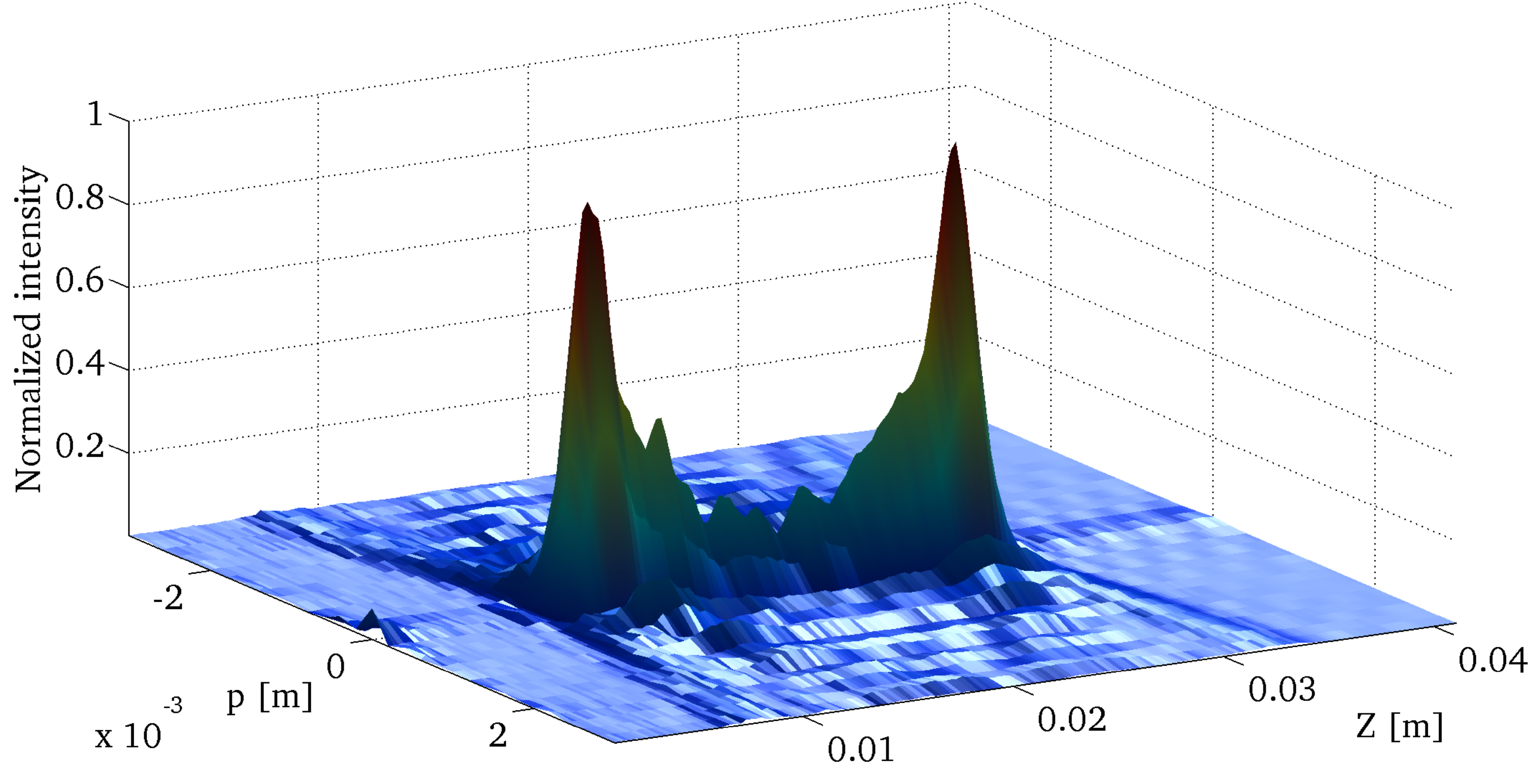}
\caption{The FW pattern, corresponding to the previous Figure, that is, to Case $2$, obtained by our
simulated experiment. The settings are the same as in the previous Figure.}
\label{fig_11}
\end{figure}

  What discussed above leads to the following question: What are the dimensions of $d$ and $\Delta_{d}$ that will permit us
to generate many different ultrasonic FW patterns with the same transducer? \ Stated in another way: What will be the
ring dimensions ($d; \ \Delta_{d}$) that best comply with a predefined set of FWs to be generated? \
To address these questions, one has first to consider the sampling process involved during the generation of the frozen waves.
This is a critical point in order to attain a proper excitation while minimizing the hardware cost.
Indeed, each transducer ring will require in practice one electronic emission channel in order to
operate.\footnote{It should be remembered that we are trying to generate localized energy spots be means of the FW, and not
dealing with ultrasound imaging.} \ For instance, the number of rings ($N_r$) and emission channels needed for the physical
realization of the patterns shown in Figures \ref{fig_8}--\ref{fig_15} is in practice of $N_r=101$; which means a considerable
amount of electronic hardware for any application. \ In conclusion, besides the required electronics,
the definition of the annular transducer itself is of key importance, and perhaps the major issue to be considered at
first in any particular application\cite{recami2011}.

  As mentioned earlier, the problem of the rings definition implies in its turn the sampling procedure. An example of
this process is shown in Figs. \ref{fig_6} and \ref{fig_7}; both corresponding to the FW envelope of Fig. \ref{fig_8}
described in Case 1 of Sec.5. \
Here the theoretical FW profiles of amplitude and phase at the aperture (i.e., at $z=0$) are sampled by adopting a
constant-step approach. The sampled values are then used as the final amplitudes and phases of the sinusoidal excitations
for each one of the transducer rings. \ The sampling process has other alternatives to be adopted, as for example a
non-constant step spacing. However, we prefer to adopt in this paper the rather simple approach of constant sampling,
leaving other possibilities for future work. In other words, once the values for $d$ and $\Delta_d$ have been assigned, we use
them for a radial {\em sampling} of the theoretical FW patterns at $z=0$: That is to say, each sample value corresponds
here to the {\em center} of each ring (i.e., the center of the segment associated to the ring width $d$).

\section{Simulation of FWs: \ Results}

\noindent To further demonstrate the possibilities offered by the ultrasound frozen waves, in this Section we present
four different simulations of ultrasonic FWs in a water-like medium, with high transverse localization [at this stage, we prefer to test simple
arbitrary patterns]. \ As mentioned at the beginning, we assume in this paper a homogeneous medium, with no
attenuation effects, operating in a linear regime; the sound speed being $c=1540\;$m/s. \ In all cases, the same
annular aperture is used, with $N_r=101$ rings endowed with width \ $d=0.3\;$mm \ and \ kerf of $\Delta_d=0.05\;$mm. \
The operating frequency is fixed at $f_0=2.5\;$MHz, which is an adequate value for use in Medicine, corresponding to a
wavelength of about $\lambda\backsimeq0.62\;$mm. \ The only parameters varied during our simulations, apart from the FW pattern
itself, are the maximum allowable distance or field-depth $L$, and the value of $N$ determining the number, $2N+1$, of Bessel
beams in Eq.(\ref{eqn_1}).

  The double Figures associated with this Section 5 (namely, Figs.8-9, 10-11, 12-13, 14-15) depict, first, the theoretical
pattern to be constructed (that is, a $3$D plot of the chosen FW); and, second, the result of the corresponding impulse
response simulation; respectively. The sampling frequency used in the IR method is $f_s=100\;$MHz.

  Let us comment about the approximate generation ---by our simulated experi-ment--- of the chosen FWs. Besides playing with
the value of $N$, which enters relation (\ref{eqn_16}), we used the tool of pushing the value of $N$ a little bit up,
for a better reconstruction of the desired ideal intensity patterns $|F(z)|^2$. \
In fact, we are still using for the aperture the same size, $R=35\;$mm, that is to say a diameter $\varnothing = 70\;$mm. \
Then, if $N^{\uparrow}$ is moderately increased, say for instance from $9$ to $12$, this trick will moderately enhance the
reconstruction of the pattern. Care should be taken, however, when the maximum range $L$ is augmented, because the pattern
would start distorting. Such an effect can be appreciated in Fig.\ref{fig_13} where the FW peak, near $z\approx 55$\;mm,
starts to loose amplitude. \

  Another issue we like to comment about changing $N$, is the variation in the size of the spot-radius of the FW given
by Eq.(\ref{eqn_9}). \ However, in the present context this has not much concern, because all spots are below $1$\;mm. \
It is also interesting to compare the long depth of field of the generated FW, with that of a gaussian
beam\footnote{The diffraction length is give by \ $z_{\text{dif}}=\sqrt{3}k_0\frac{\Delta\rho_0^2}{2}$. \ See pp.8-9 of
Ref.\cite{hugo2008}.}, which in the best case with the current setting would be $z_{\text{dif}}\backsimeq 4$\;mm.

  Because of the computer time employed by the IR simulations, the details in the corresponding spatial grids ($\rho,z$)
had to be reduced w.r.t. the ideal patterns. Then, suitable intervals of $\Delta_z=0.25\;$mm \ and \ $\Delta_{\rho}=0.15\;$mm
have been selected for the simulated plots. \ Also, due to the adoption of colors for the matlab graphs (absent, however, in the
version printed on paper), an effect was added to enhance the visibility of the smaller details of the patterns.

\subsection{Case 1}
\noindent As a first choice, the FW to be reproduced consists in two step-functions with different amplitudes. The maximum
distance selected for this case is $L=30\;$mm, \ the number of the Bessel beams to be superposed being $2N+1=25$ . The
the spot size is approximately $2\Delta\rho_1 \backsimeq 0.72$\;mm. The corresponding envelope function $F_1(z)$ is
therefore
\begin{equation}\label{eqn_18}
F_1(z)= \begin{cases}
      0.5   \ &\text{for $l_1 \leq z \leq l_2$}\\
      1 \ &\text{for $l_3 \leq z \leq l_4$}\\
    \end{cases}
\end{equation}

\noindent with $l_1=L/6$, \ $l_2=2L/6$, \ $l_3=3L/6$ \ and \ $l_1=5L/6$; \ the field being zero elsewhere.

  The theoretical pattern is shown in Fig.\ref{fig_8}, while the IR simulation is shown in Fig.\ref{fig_9}.
Notice how the peaks and the valleys of the ideal FW (which corresponds to $N=12$) are clearly emulated by the
results of our simulated experiment depicted in Fig.\ref{fig_9}. Only in the region near $z\approx 5\;$mm the
simulation begins to deviate from the theoretical behavior.

\begin{figure}[t]  
\centering
\includegraphics[width=3.5in]{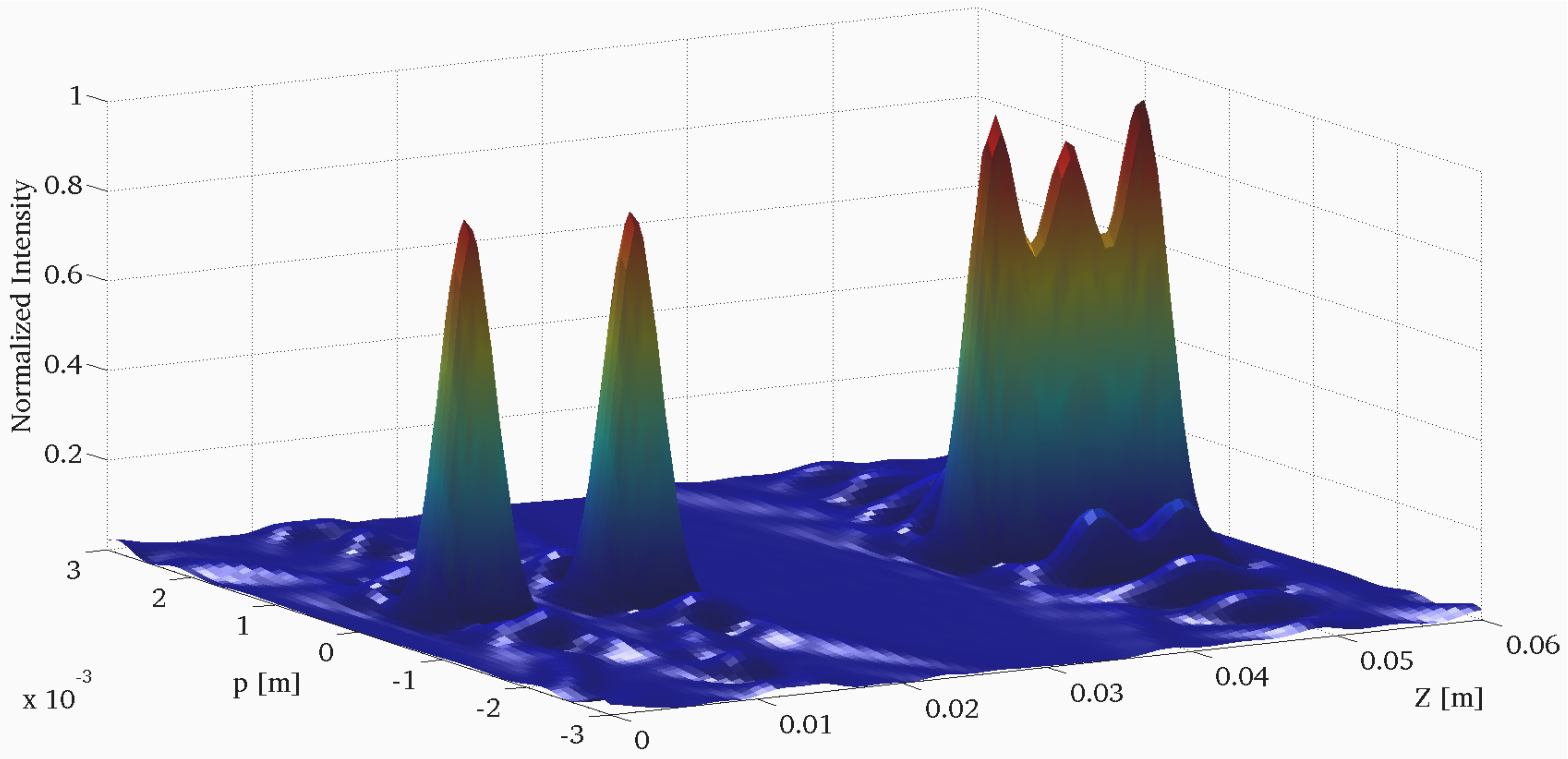}
\caption{The theoretical FW pattern chosen in Case 3 (see the text). Settings: $N=11$; \ $R=35\;$mm; \ $d=0.3\;$mm; \
$\Delta_d=0.05\;$mm; \
$N_r=101$; \ $L=60\;$mm, \ and  $f_0=2.5\;$MHz.}
\label{fig_12}
\end{figure}

\begin{figure}[t]  
\centering
\vspace{-2.2mm}
\includegraphics[width=3.5in]{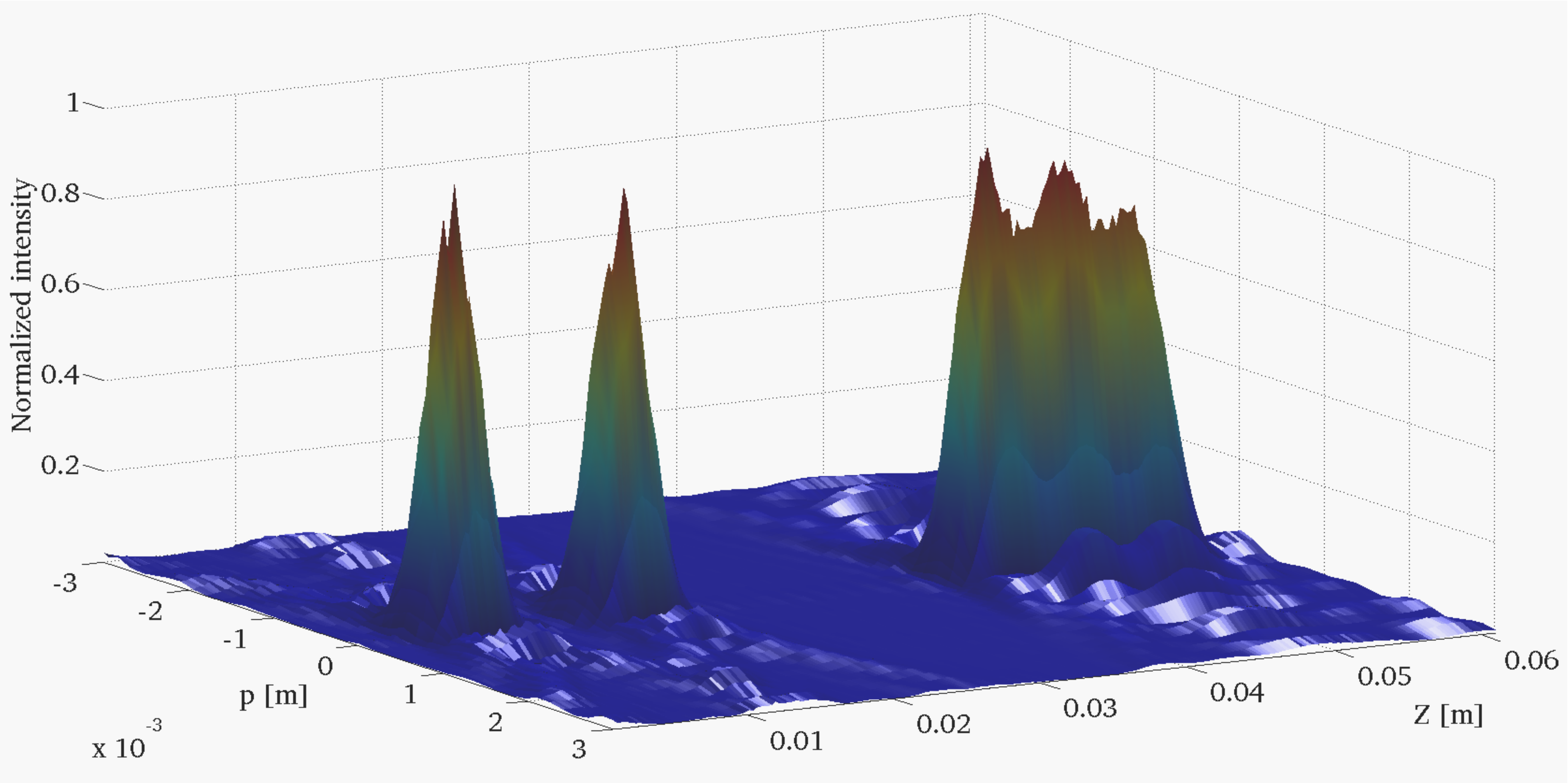}
\caption{The FW pattern, corresponding to the previous Figure, that is, to Case 3, obtained by simulated experiment.
The settings are as above.}
\label{fig_13}
\end{figure}

\subsection{Case 2}
\noindent The second pattern, selected for the FW to be created, consists in a concave-shaped region, whose envelope
function is therefore
\begin{equation}\label{eqn_19}
F_2(z)= 1 + 2.5\frac{(z-l_1)(z-l_2)}{(l_2-l_1)^2}   \ \ \ \text{for $l_1 \leq z \leq l_2$}
\end{equation}

\noindent with $l_1=2L/8$ and $l_2=6L/8$. The maximum distance adopted in this case is $L=40\;$mm, and the number of the
superposed Bessel beams is again $2N+1=25$. This corresponds now to a transverse spot of $2\Delta\rho_2 \backsimeq 0.82$\;mm.

  Figures \ref{fig_10} and \ref{fig_11} show the corresponding theoretical and simulated ultrasonic field, respectively.
Again, a good match is obtained between the theoretical and the simulated FW, notwithstanding the rather strange look of the
pattern.

\subsection{Case 3}
\noindent This case corresponds to the field depicted in Figures \ref{fig_12} and \ref{fig_13}. Here we purpose to create
two small convex, or peaked, field regions, followed by a constant-field finite region (a ``step"); the field being assumed to be
zero elsewhere. \ The envelope function
employed this time is therefore given by (the field, let us repeat, being otherwise {\em zero\/}):
\begin{equation}\label{eqn_20}
F_3(z)= \begin{cases}
      -4\frac{(z-l_1)(z-l_2)}{(l_2-l_1)^2}  \ \ &\text{for $l_1 \leq z \leq l_2$}\\
      -4\frac{(z-l_5)(z-l_6)}{(l_6-l_5)^2}  \ \ &\text{for $l_3 \leq z \leq l_4$}\\
      \ \ \ \ \ \ \ \ \ 1           \ \ &\text{for $l_5 \leq z \leq l_6$}\\
        \end{cases}
\end{equation}

\noindent with $l_1=0.5L/8$, \ $l_2=1.5L/8$, \ $l_3=2L/8$, \ $l_4=3L/8$, \ $l_5=5.5L/8$ \ and \ $l_6=7.5L/8$. The
obtained spot-size is $2\Delta\rho_3 \backsimeq 0.98$\;mm.

  As we stated before, we can see the effect on the second peaked region produced by augmenting the number of the beams, which
from Eq.\ref{eqn_16} should be now $N'=11$ instead of $N\simeq 7$. Also, the field-depth is now larger, that is, $L=60\;$mm.

\begin{figure}[t]  
\centering
\includegraphics[width=3.5in]{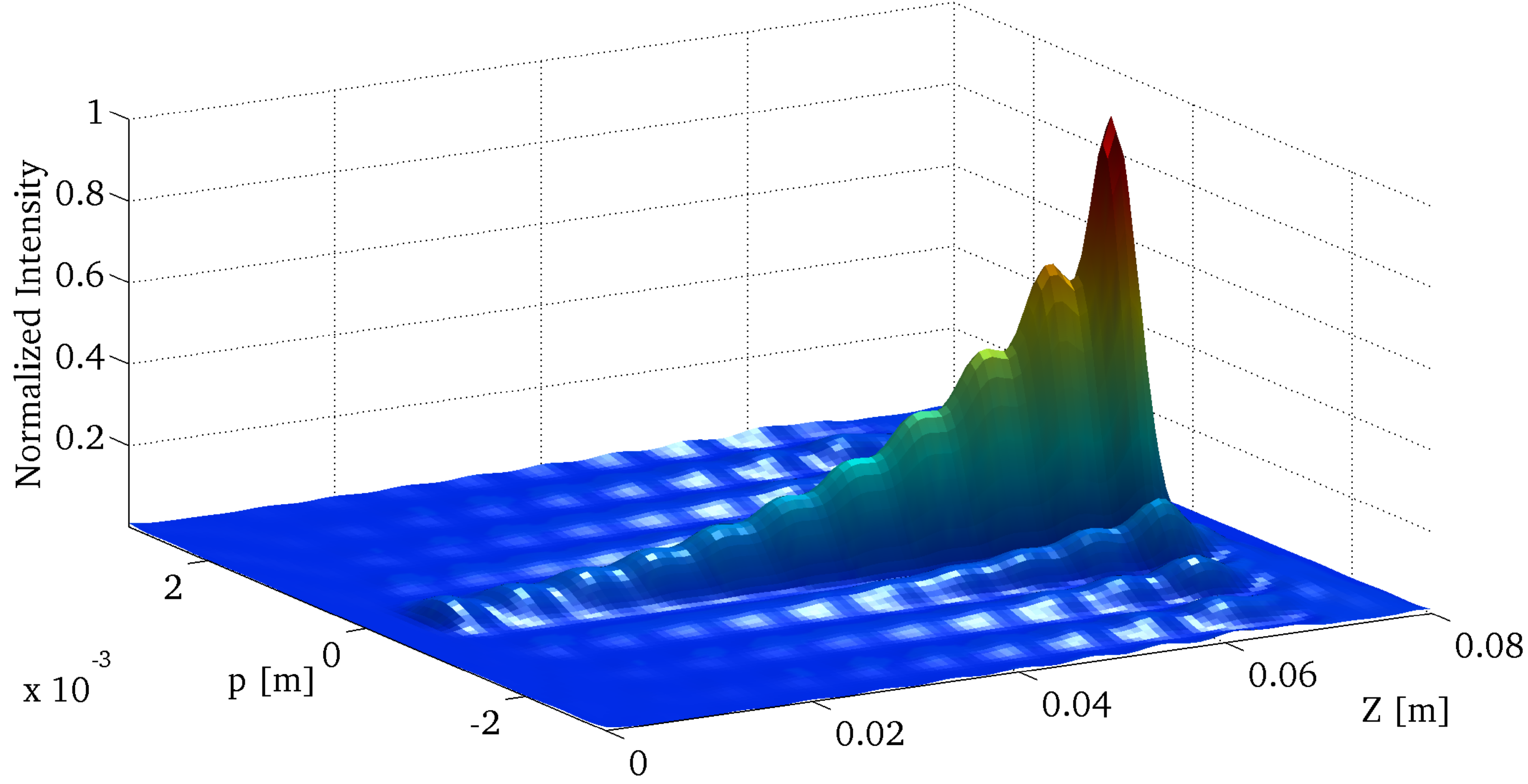}
\caption{The theoretical FW pattern of Case 4. Settings: $N=8$; \ $R=35\;$mm; \ $d=0.3\;$mm; \ $\Delta_d=0.05\;$mm; \ $N_r=101$; \
$L=80\;$mm, \and \ $f_0=2.5\;$MHz.}
\label{fig_14}
\end{figure}

\begin{figure}[t]  
\centering
\includegraphics[width=3.5in]{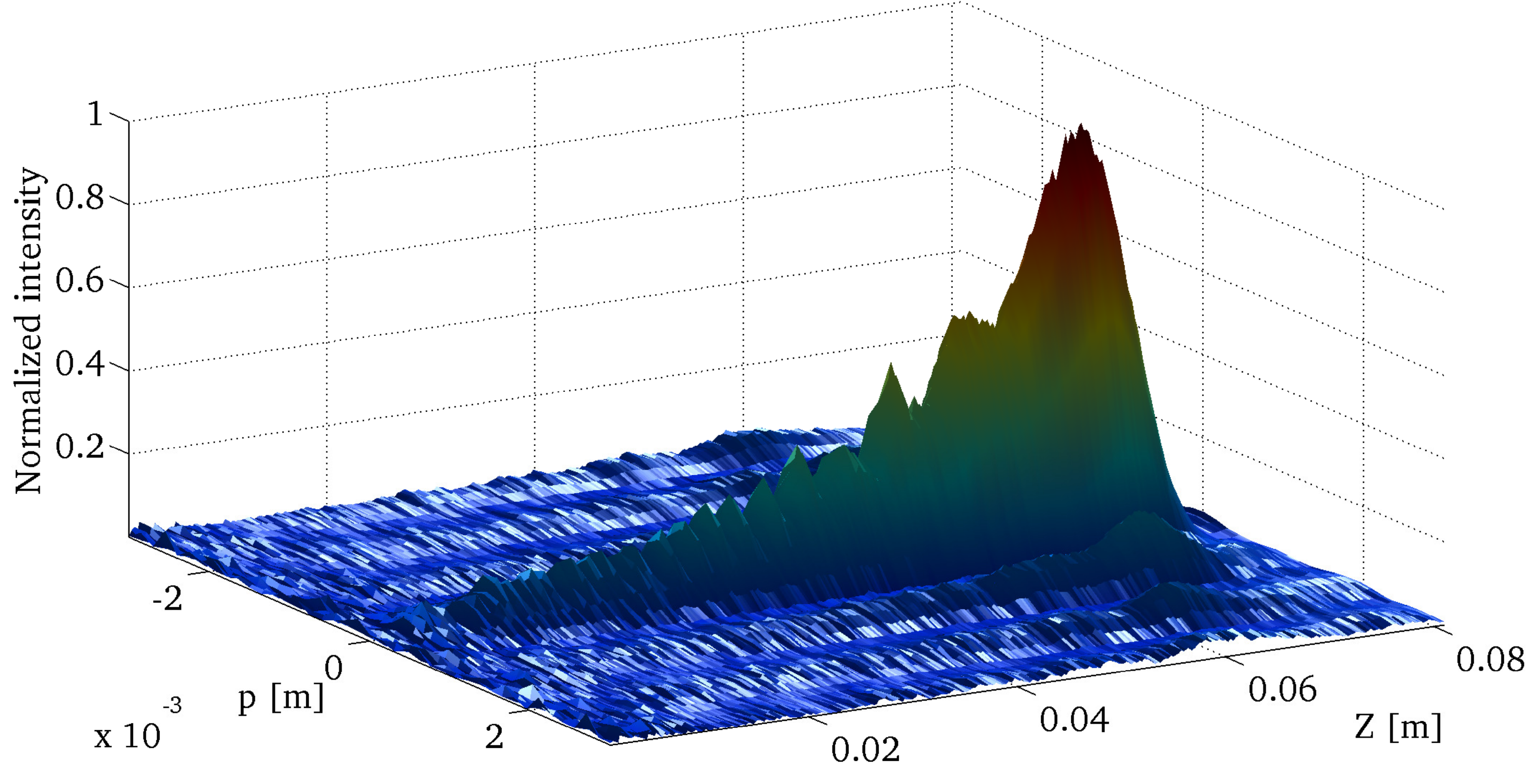}
\caption{The FW pattern, corresponding to the previous Figure, that is, to Case 4, obtained by our
simulated experiment. The settings are as above.}
\label{fig_15}
\end{figure}

\subsection{Case 4}
\noindent The last example corresponds to an exponentially growing FW, with a maximum distance now even larger, of $L=80$\;mm.

  This is a rather important example, since it is directly related to the generation of a FW in an attenuating medium. The
associated envelope function $F_4(z)$ is now
\begin{equation}
\label{eqn_21}
F_4(z)= \frac{e^{\frac{3z}{L}}}{5}\ \ \ \ \ \text{for $l_1 \leq z \leq l_2$}\\
\end{equation}

\noindent with $l_1=0.5/8$ and $l_2=7.5L/8$.

  In order to enhance the fidelity in the reconstruction of the ideal $F_4(z)$ envelope, and at the same time to reduce
the effects produced by using augmented $N^{\uparrow}$ and $L$ values, the number of the superposed Bessel beams
is now taken to be $17$ \ ($N=8$). \ The transverse spot-size this time results to be $2\Delta\rho_4 \backsimeq 1.36$\,mm.

  Notice how the ultrasonic field in Fig.\ref{fig_15} simulates in a pretty good way the theoretical $3$D plot of the FW
in \ref{fig_14}. Good results of this type encourage to generate frozen waves,
even in attenuating media.

\section{Some Conclusions and Prospects}

\noindent In this paper we have shown how adequate superpositions of zero order Bessel beams can be used for
constructing ultrasonic wavefields with a {\em static} envelope, within which only the carrier wave propagates;
for simplicity, we have here assumed a water-like medium, disregarding attenuation. \ Indeed, we have demonstrated by
simulated experiments, via the impulse response method, that suitable sets of annular apertures can
produce ultrasonic fields ``at rest'': which have been called ultrasonic {\em frozen waves} (FW) by us. Such FWs belong
to the realm of the so-called localized waves (LW), or Non-diffracting Waves, which are soliton-like solutions to the
ordinary {\em linear} wave equation which since long time have been shown theoretically and experimentally to be endowed with
peak-velocities $V$ ranging  from $0$ to $\infty$ \cite{hugo2008, recami2009}.

  The FWs, among the sub-luminal (or rather sub-sonic, in our case) LWs\cite{zamboni2008}, are the ones
associated\cite{zamboni2004,zamboni2005} with $V=0$.

  An important characteristic of the FWs is that they can be constructed, within a prefixed spatial interval
$0 \leq z \leq L$, with any desired longitudinal intensity shape, maintaining at the same time a high transverse localization.
This has been verified through the size of the intensity {\em spots} of the generated FWs, which have almost reached the
diffraction limit $\lambda$. All this can be monitored by appropriate selection of the parameter $Q$.\\

  We also pointed out how  ---when adopting for instance sets of annular radiators--- a proper generation of
ultrasonic FWs with a given frequency requires careful attention to the dimensioning of the annular transducers. \
Namely, we have discussed how the transducer radii ($R$), the operating frequency ($f_0$),
and the ring dimensions (width$=d$ and kerf$=\Delta_d$) affect the generated FW fields. \ We have shown, in particular,
how a rise in frequency allows improving the fidelity of the generated FWs while
keeping the emitter radii $R$ unchanged, and just working on (augmenting) the parameter $N$; even if
such an increase of $N$ will obviously impose sharper requirements on the emitter ring-elements.

  Associated to the problem of defining the ring sizes, another important issue we have pointed out is the required
sampling process of the FW patterns at the aperture: Something that one must bear in mind when designing the annular arrays
for the considered FW.  \  To address this point, we explored in this article a simple constant-step sampling approach,
leaving other alternatives for future work, in which we also intend to consider pulsed exitations\cite{lu1992b,castellanos2010}
as a possible way for surmounting the known side-lobe problem, met with the use of Bessel beams for focusing ultrasonic
energy in CW (continuos wave) mode.

  As a merely complementary issue, we discussed incidentally the effect of raising even more the value of
$N^{\uparrow}$ [instead of sticking to the value yielded by Eq. (\ref{eqn_16})] for getting in more detail
the ideal envelope $F(z)$ (see Section 2), neglecting for the moment the transverse-spot size changes.

  To spend a few more words on the practical realization of acoustic FWs, let us recall that significant efforts
from the point of view of the hardware could be requested. We feel that the main difficulty is linked however
with proper choice, and development, of the ultrasonic transducers themselves: Namely, of transducers with
suitable dimensions and $\lambda/4$ adaptation layer(s), as well as with a low parasitic inter-element {\em cross-coupling}
(e.g., lower than $-35$\,dB).  According to us, with one such transducer in the hands, acoustic FWs can be experimentally
tested. Further work and attention will have to be devoted, in any case, to ease up the mentioned technological complexities,
and costs, which in principle can be involved in the construction of the suitable CW driving electronics for
the required annular array. Without forgetting that we confined ourselves, for obvious reasons, to study the region located
between the aperture and the propagating medium, disregarding at this stage {\em possible} issues referring to the
transfer functions associated with each one of the annular electromechanical elements: Something that could have indeed
its role when aiming at an efficient implementation with pulsed HV exitations.\  In the present work, the latter problems
do not show up, in practice, since we refer to operation in the CW regime; so that all the ring elements, plus the channels of
the electronic front-end (LECOUER tm), are tuned at the resonant frequency of the array piezoelectric dye.

  Many applications of the localized FWs in various sectors of science are possible. In the case of
ultrasonic FWs, let us mention for example new kinds of acoustic tweezers, bistouries, and other possible medical apparatus
for the treatment of affected tissues. With respect the last point, we believe that FWs can offer for example
an alternative to HIFU\footnote{High-Intensity Focused Ultrasound.}, since they allow an extended control and modeling
of the longitudinal field envelope. A further advantage of acoustic FWs is  that they don't have to deal
with the drawback, of relative small treatment volume versus relatively large access window, usually
met\cite{rossing2007} by the HIFU techniques.

\section{Acknowledgments}
\addcontentsline{toc}{section}{Acknowledgment}
\noindent The authors thank Dr. Antonio Ramos, from CSIC, Spain, and Dr. Jos\'{e} J. Lunazzi, from
UNICAMP, Brazil, for their kind help and interest. They are moreover grateful to the Topical Editor
for kind attention, and to the anonymous Referees for useful comments. \ This work was supported by FAPESP, Brazil
(as well as, partially, by CAPES and CNPq, Brazil, and INFN, Italy).

\ifCLASSOPTIONcaptionsoff
  \newpage
\fi

\end{document}